\documentclass[9pt,twocolumn,twoside]{osajnl}

\journal{ol} 

\setboolean{shortarticle}{true}
\DeclareMathOperator{\Tr}{Tr}
\DeclareMathOperator{\Det}{Det}

\title{Coherence-induced Polarization Effects in  Vector Vortex Beams}

\author[1]{Stuti Joshi}
\author[1,*]{Saba N Khan}
\author[1]{Manisha}
\author[1]{P Senthilkumaran}
\author[1]{Bhaskar Kanseri}

\affil[1]{Department of Physics, Indian Institute of Technology Delhi, Hauz Khas, New Delhi 110016, India}

\affil[*]{Corresponding author: sabakhan@iitd.ac.in}

\begin{abstract}
We demonstrate theoretically and experimentally coherence-induced polarization changes in generic and higher-order vector vortex beams with polarization singularity. The prominent depolarization on decreasing transverse correlation-width in focused partially coherent vector vortex beam provides a means to shape the intensity profile and the degree of polarization (DOP) while preserving the polarization distribution. The intensity variation and DOP-dip are found to be dependent on the polarization singularity index of the beam. Our results may provide an additional degree of freedom in the myriad of applications presently projected with various types of vector vortex beams.
\end{abstract}

\setboolean{displaycopyright}{true}

\begin{document}

\maketitle
Polarization and spatial coherence were regarded as mutually independent properties until a unified theoretical framework based on field correlations was developed \cite{wolf2007introduction}. Degree of polarization (DOP), which has been considered as an intrinsic property of the electromagnetic beam, thereafter found ways to be tuned \cite{wolf2003correlation}. In far-field propagation, the on-axis DOP of electromagnetic beams is coarsely tuned by introducing distinct transverse correlation-widths ($\delta_{xx}\ne \delta_{yy}\ne\delta_{xy}$) \cite{james1994change, korotkova2005changes}. However, one hardly realizes any change in on-axis DOP for equal correlation-widths ($\delta_{xx}= \delta_{yy}$). Recently, the ways to modulate the DOP in the beam cross-section (transverse plane) using 2$f$ and 4$f$ lens systems \cite{zhao2018controlling,zhao2018degree} were reported. Contrary to maximum DOP at the centre of the scalar beams, the vector vortex beams always have a zero DOP at the central-core because of the absence of polarized electric-fields. Interestingly, on embedding an additional phase vortex over a generic vector-vortex beam an anti-depolarization effect around the central core of the beam has been observed \cite{zeng2020partially}. Due to the interplay of coherence and polarization, the modulation in DOP on changing the correlation in fluctuating fields is expected in a standard Gaussian, Laguerre-Gaussian, Hermite-Gaussian, Bessel-like beams, etc. \cite{zhao2018controlling,liu2019review,ji2009changes,korotkova2005changes,kanseri2013optical}. In fact, beam-shaping  by controlling field correlation between two transverse points (spatial coherence) in partially coherent vortex beams has been established \cite{liu2019review}. The fluctuating fields also manifest the change in the state of polarization (SOP) \cite{korotkova2005changes}.
Hence, vector vortex fields are appropriate field-structures for investigating the coherence-induced polarization changes. 

Vector vortex fields are embedded with polarization singularities in the core \cite{Senthil}. The SOP distribution around the singular point can be- radial, azimuthal, spider-web, flower-like, etc \cite{freund2001polarization}. Vector field singularity is characterized by Poincar\'e-Hopf index (PHI), which defines the strength of azimuth gradient \cite {dennis2009singular}. The PHI is given by $\eta = \frac{1}{2\pi}\oint \nabla \gamma \cdot dl$ evaluated around the singularity. Here, $\gamma$ is the orientation of linear polarization (azimuth). This expression looks similar to the one used for phase singularity determining topological charge. The line integral evaluates azimuth gradient for PHI and  phase gradient for topological charge \cite{Senthil}. Although there are studies  on polarization singular beams of fully coherent fields \cite{Senthil,milione2011higher,dennis2009singular,khan2020perturbation, freund2001polarization}, for partially coherent VVBs the studies are limited to preliminary reports \cite{guo2011intensity, wang2012experimental,dong2012effect,gbur2020singularities,chen2014generation}.
 
In this letter, we report coherence-induced polarization changes in generic as well as higher-order vector vortex beams (VVB) endowed with singularities. We show that the transverse-DOP distribution of focused partially coherent vector vortex beam (PCVVB) depends strongly upon the spatial correlation of the input beam while the SOP distribution remains invariant. This means that although $S_1^2+S_2^2+S_3^2<1$, the Stokes parameters $S_1, S_2$ and $S_3$ are such that the SOP distribution is preserved. As expected, self-shaping of the beam is also observed on varying the correlation between two transverse points in the source plane \cite{guo2011intensity,wang2012experimental}. The variation in the intensity profile and the DOP-dip are also found to be dependent on the associated polarization singularity index ($\eta$) of the beam. Such a source with partial coherence and partial polarization is useful in long-distance communication by reducing the scintillation index in atmospheric turbulence \cite{korotkova2008scintillation}. Besides, the invariant polarization distribution with good control of transverse DOP profile can provide an additional degree of freedom for such VVBs in classical \cite{Senthil,dennis2009singular} and quantum domain \cite{ndagano2017characterizing}.\begin{figure} [ht!]
\centering
\includegraphics[width=0.95
\linewidth]{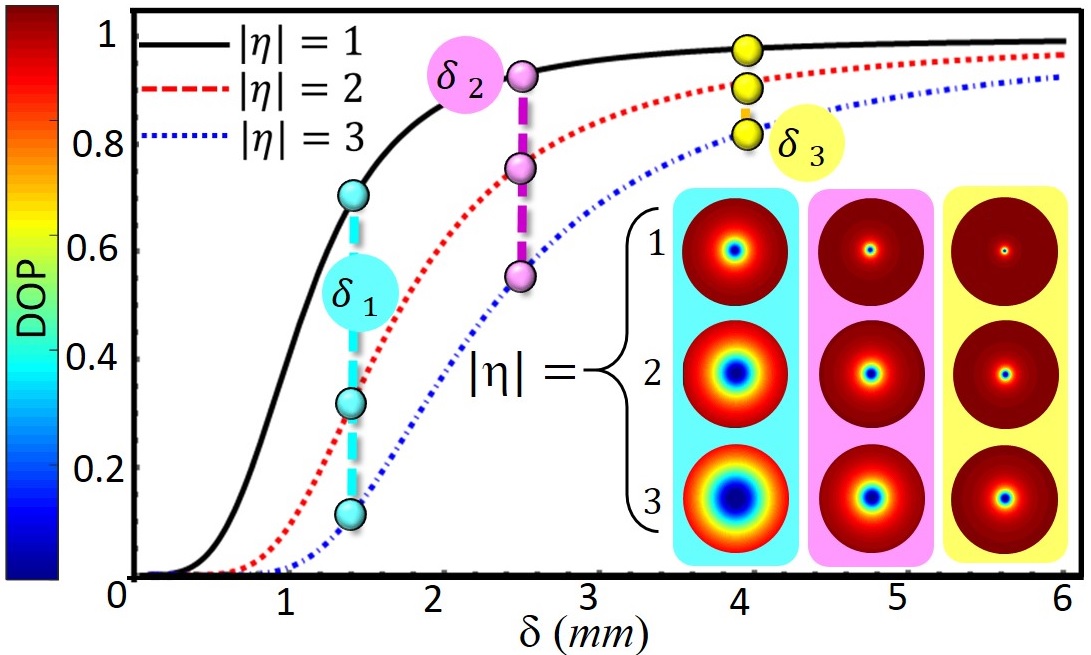}
\caption{Simulation: Dependency of DOP on correlation-width ($\delta$) of focused ($f=300 \thinspace mm$) PCVVBs at (0.4 $mm$,0) for various $\eta's$. The chosen point corresponds to maximum DOP on the edge of a highly coherent PCVVB of beam-waist $\sigma\sim$ 2.4 $mm$. Density plots of DOP for three values of $\delta$ are shown.}
\label{fig:DOP}
\end{figure}
\begin{figure*} [ht!]
\centering
\includegraphics[width=0.9\linewidth]{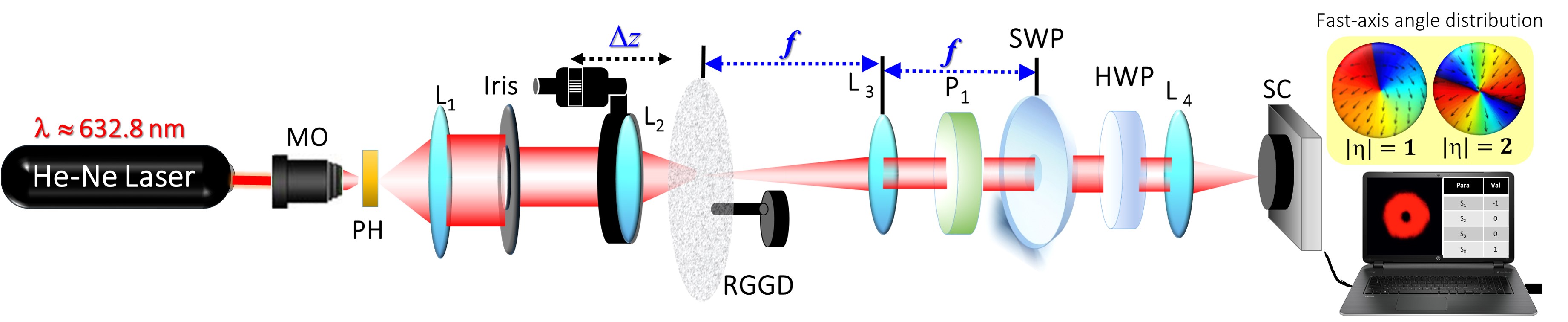}
\caption{Experiment: Investigation of coherence-induced polarization changes in PCVVBs. $L_1$, $L_2$, $L_3$ and $L_4$: lens; MO: microscope objective; PH: pinhole; P: polarizer; SWP: spatially-varying wave plate; HWP: half-wave plate; RGGD: rotating ground glass diffuser and SC: Stokes camera. Beam-size at RGGD is measured by beam profiler. Inset shows fast-axis angle distribution on SWP.}
\label{fig:experimental}
\end{figure*}

Electric field of a coherent VVB is superposition of orthogonal polarized vortex ($+l$) and anti-vortex ($-l$) states \cite{Senthil}, as
\textcolor{black}{
\begin{eqnarray}
\textbf{E}(r,\phi)= 
\resizebox{0.76\hsize}{!}{$\frac{i}{2}[{\pm}r^{|l|} e^{{\pm}il\phi}(\hat{\text{x}}-i\hat{\text{y}})- r^{|l|}e^{{\mp}il\phi}(\hat{\text{x}}+i\hat{\text{y}})],\thinspace (\text{I \& II})$}\nonumber\\
=\resizebox{0.76\hsize}{!}{$\frac{1}{2}[r^{|l|} e^{{\pm}il\phi}(\hat{\text{x}}-i\hat{\text{y}}){\pm} r^{|l|}e^{{\mp}il\phi}(\hat{\text{x}}+i\hat{\text{y}})],\thinspace (\text{III \& IV})$}
\end{eqnarray}} where $r^2=x^2+y^2$. $l$ and $\phi$ are topological charge and azimuthal angle, respectively. \textcolor{black}{The PHI of the resulting beam is such that $|\eta|=|l|$.} For a given $\eta$, there are two pairs of orthogonal states (Type I and Type III; Type II and Type IV) that are intensity degenerate \cite{ram2017probing}. Following the unified theory \cite{wolf2007introduction}, for a statistically stationary, quasi-monochromatic PCVVB, the coherence properties in space-frequency domain are characterized by a $2{\times}2$ cross-spectral density (CSD) matrix $\textbf{W}(r_1,\phi_1;r_2,\phi_2)$. For a PCVVB generated by a Gaussian Schell-model source of beam waist $\sigma$ and spatial correlation-width $\delta$, in the cylindrical coordinate system, the elements of its CSD matrix (Type I PCVVB) in the source plane are
{\begin{eqnarray}
\resizebox{.9\hsize}{!}{$W_{0xx}(r_1,\phi_1;r_2,\phi_2)=\frac{(r_1r_2)^{|l|}}{(2\sigma)^{2|l|}}\sin(l\phi_1)\sin(l\phi_2)\mu(\textbf{r}_1,\textbf{r}_2),$}\\
\resizebox{.9\hsize}{!}{$W_{0yy}(r_1,\phi_1;r_2,\phi_2)=\frac{(r_1r_2)^{|l|}}{(2\sigma)^{2|l|}}\cos(l\phi_1)\cos(l\phi_2)\mu(\textbf{r}_1,\textbf{r}_2),$}\\
\resizebox{.9\hsize}{!}{$W_{0xy}(r_1,\phi_1;r_2,\phi_2)=-\frac{(r_1r_2)^{|l|}}{(2\sigma)^{2|l|}}\sin(l\phi_1)\cos(l\phi_2)\mu(\textbf{r}_1,\textbf{r}_2),$}\\
\resizebox{.9\hsize}{!}{$W_{0yx}(r_1,\phi_1;r_2,\phi_2)=-\frac{(r_1r_2)^{|l|}}{(2\sigma)^{2|l|}}\cos(l\phi_1)\sin(l\phi_2)\mu(\textbf{r}_1,\textbf{r}_2),$}
\end{eqnarray}
where $\mu(\textbf{r}_1,\textbf{r}_2)=\exp\left(-\frac{r_1^2+r_2^2}{4\sigma^2}\right)
\exp\left(-\frac{r_1^2+r_2^2-2r_1 r_2\cos(\phi_1-\phi_2)}{2\delta^2}\right)$ and $\textbf{r}_1(r_1,\phi_1)$; $\textbf{r}_2(r_2,\phi_2)$ are two points in the source plane.} Noteworthy, for a quasi-monochromatic field, the elements of both beam coherence polarization matrix and the CSD matrix possess identical values \cite{kanseri2010experimental}.
\begin{figure} [ht!]
\centering
\includegraphics[width=1.0
\linewidth]{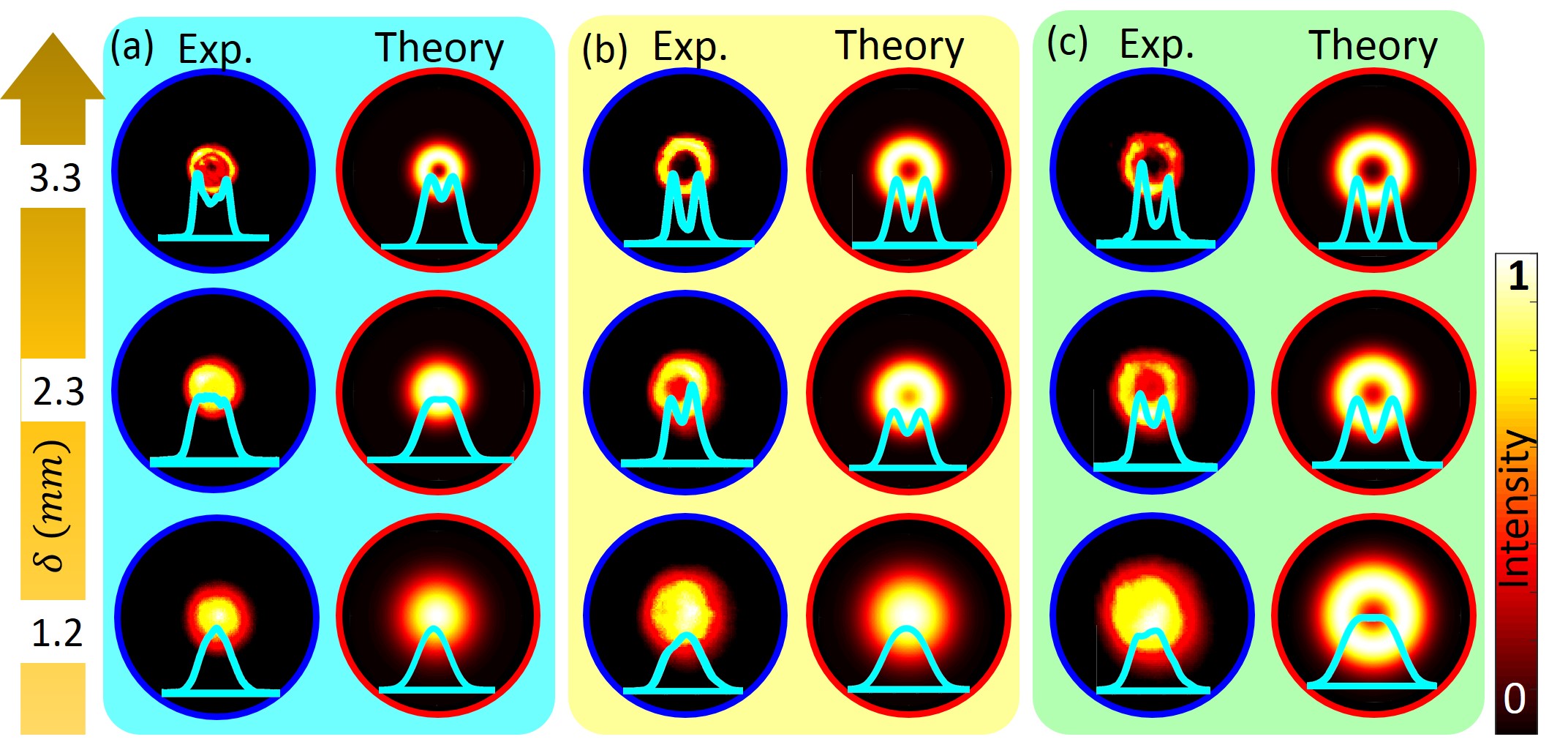}
\caption{Normalized intensity distributions of PCVVBs for different values of $\delta$; (a) $|\eta|=1$, (b) $|\eta|=2$ and (c) $|\eta|=3$. Beam shaping of PCVVB can be seen from line-profiles.}
\label{fig:3}
\end{figure}
The elements of the CSD matrix propagating through an ABCD optical system can be evaluated by using generalized Collins formula \cite{lin2002tensor}
\begin{equation}
\begin{split}
W_{\alpha\beta}(\boldsymbol\rho_1,\boldsymbol\rho_2)=\frac{1}{\lambda^{2}B^2}\int_{0}^{2\pi}\int_{0}^{2\pi}\int_{0}^{\infty}\int_{0}^{\infty}r_1r_2{dr_1}{dr_2}{d\phi_1}{d\phi_2}\\
{\times}W_{0\alpha\beta}(r_1,\phi_1;r_2,\phi_2)\exp\left[\frac{ikD}{2B}(\rho_2^2-\rho_1^2)-\frac{ikA}{2B}(r^2_1-r^2_2)\right]\\
\times\exp\left[\frac{ik}{B}\left(r_1\rho_1\cos(\theta_1-\phi_1)-r_2\rho_2\cos(\theta_2-\phi_2)\right)\right],
\end{split}
\end{equation}
where $k=2\pi/{\lambda}$, $\lambda$ being the wavelength. The subscripts $\alpha=x,{\thinspace}y;{\thinspace}\beta=x,{\thinspace}y$ and  $\boldsymbol{\rho}_1\thinspace(\rho_1,\theta_1)$; $\boldsymbol{\rho}_2\thinspace(\rho_2,\theta_2)$ represent the coordinates of two transverse points in PCVVBs at the  observation plane. The elements of transfer matrix ABCD for far-field propagation are, $A$ = $D$ = 0, $B$ = $f$
and $C=-1/f$ \cite{wang2012experimental}.
The intensity distribution and the DOP can be calculated using \cite{wolf2007introduction}
\begin{eqnarray}
  I(\boldsymbol\rho)=W_{xx}(\boldsymbol\rho,\boldsymbol  \rho)+W_{yy}(\boldsymbol\rho,\boldsymbol\rho),\\
   P(\boldsymbol\rho)=\frac{I_p(\boldsymbol\rho)}{I(\boldsymbol\rho)}=\sqrt{1-\frac{4{\Det}[\textbf{W}(\boldsymbol\rho,\boldsymbol\rho)]}{({\Tr}[\textbf{W}(\boldsymbol\rho,\boldsymbol\rho)])^2}},
\end{eqnarray}    
where Det and Tr denote the determinant and trace of the CSD matrix respectively. The Stokes parameters are connected with the elements of the CSD matrix as \cite{wolf2007introduction}
\begin{equation}
 \left.
    \begin{array}{ll}
S_0(\boldsymbol\rho)=W_{xx}(\boldsymbol\rho,\boldsymbol\rho)+W_{yy}(\boldsymbol\rho,\boldsymbol\rho), \\
S_1(\boldsymbol\rho)=W_{xx}(\boldsymbol\rho,\boldsymbol\rho)-W_{yy}(\boldsymbol\rho,\boldsymbol\rho), \\
S_2(\boldsymbol\rho)=W_{xy}(\boldsymbol\rho,\boldsymbol\rho)+W_{yx}(\boldsymbol\rho,\boldsymbol\rho), \\
S_3(\boldsymbol\rho)=i(W_{yx}(\boldsymbol\rho,\boldsymbol\rho)-W_{xy}(\boldsymbol\rho,\boldsymbol\rho)).
    \end{array}
\right \} 
\end{equation}
We have numerically solved Eqs. (6)-(9) to determine intensity, DOP and SOP distributions. 

Figure \ref{fig:DOP} shows the change in DOP of a focused PCVVB as a function of correlation-width ($\delta$). Three different VVBs with $|\eta|= 1,  2$ and $3$ are considered in this study. The DOP deteriorates more rapidly for a higher PHI-PCVVB for less correlated fields. For instance, $\delta_{1}$ = 1.4 mm; DOP$_{(\eta = 1)}$= 0.7, DOP$_{(\eta = 2)}$= 0.3, DOP$_{(\eta = 3)}$= 0.1.  However, for large $\delta$, the DOP becomes nearly independent of PHI ($\eta$) of PCVVBs. This can be easily perceived from the spreading of the dark-core region in the density plots of DOP of Fig. \ref{fig:DOP} (inset). The dependence of DOP on $\delta$ and $\eta$ can be explained as follows: DOP is given by the ratio of the intensity of polarized part to the total intensity of the beam at a fixed space-point, while $\delta$ refers to the strength of spatial correlation of the fluctuating fields at two-space points \cite{born2013principles}. Reduction in $\delta$ implies a decrease in spatial correlation at the source plane (i.e., degree of coherence, DOC), which on propagation results in decreasing the polarized contribution at the observation plane. Hence, the DOP decreases at the focal plane  \cite{chen2014generation}. \textcolor{black}{Such a DOP-modulation even for an isotropic source {($\delta_{xx}=\delta_{yy}=\delta_{xy}$)} is due to the inhomogeneous SOP distribution of the beam at the source plane \cite{james1994change}}. Higher-order VVBs are more unstable and a small perturbation splits them into unit-index VVBs that propagate independently \cite{khan2020perturbation}. Therefore, the field correlation of higher-order PCVVBs depreciates more quickly reducing the DOP further. The DOP of PCVVBs at observation plane depends on both DOP and DOC at the source plane. Notably, the DOP at the source plane is always unity at all the points and it does not change on propagation for a completely correlated-field. But for less correlated fields (i.e., PCVVBs) the numerical results predict that the fully polarized PCVVBs (source-plane) become partially polarized at the focal plane \cite{chen2014experimental}.  

Experimental setup to synthesize PCVVBs with controllable spatial $\delta$ is shown in Fig. \ref{fig:experimental}. The cylindrically polarized vortex beams can be generated using spatially varying wave plate (SWP, Model: WPV10L-633, Thorlabs), which is made of half wave plate segments whose fast axis is spatially varying \cite{Senthil,ram2017probing}. Collimated linearly polarized beam of He-Ne laser ($\lambda=632.8{\thinspace}nm$) illuminates rotating ground glass diffuser (RGGD) through lens $L_2$ to obtain an incoherent light source. The light field was made partially coherent at SWP by placing it at the back focal plane of lens $L_3$ with RGGD at the front focal plane \cite{foley1991effect}. The SWP embeds the polarization distribution while maintaining unit DOP at this transverse plane. The generated PCVVB was focused by lens $L_4$ ($f=300{\thinspace}mm$) to investigate its far-field properties. Stokes Camera (SALSA, Bossa Nova Technologies) was used to record the respective Stokes parameters $(S_{i}(\boldsymbol\rho);i = 0,1,2,3)$ from which the intensity profiles $(S_{0}(\boldsymbol\rho))$ and DOP-distributions $\left(\frac{\sqrt{S_{1}^2+S_{2}^2+S_{3}^2}}{S_{0}}\right)$ of the PCVVBs are obtained. The beam-size at RGGD was varied by translating lens $L_2$ to vary the correlation-widths \cite{foley1991effect}. For our input beam-waist $\sim$ 2.4 $mm$, $\delta$ ranging from $4.4{\thinspace}mm$ to $1.2{\thinspace}mm$ were obtained.

\begin{figure} [ht!]
\centering
\includegraphics[width=1.0
\linewidth]{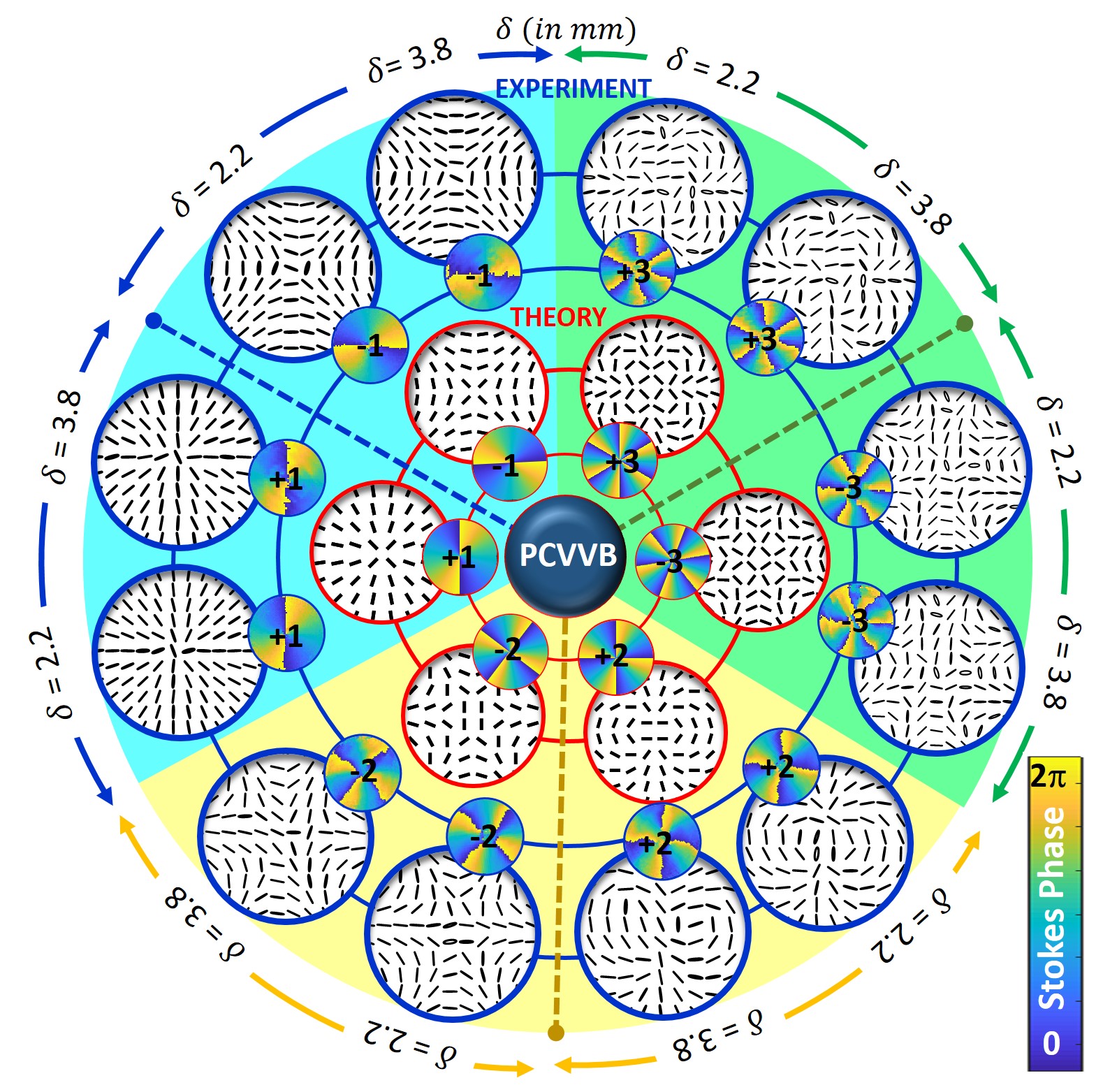}
\caption{Simulated (inner-circle) and experimental (outer-circle) Stokes phase, SOP distribution of PCVVBs for two values of $\delta$ (2.2 $\thinspace mm \thinspace \text{and} \thinspace 3.8 \thinspace mm$). Cyan, yellow and green backgrounds carry four-types of beams with $|\eta|=1$, $2$ and $3$; respectively.}
\label{fig:4}
\end{figure}

Theoretical and experimental focused intensity profiles of various PCVVBs are shown in Fig. \ref{fig:3}. The intensity profile of all four types of PCVVB of a particular $|\eta|$ are degenerate. The profile gradually transforms from donut to flat-top and finally to Gaussian distribution with a decrease in correlation-width ($\delta$) in the source plane. The flat-top intensity profiles, which have many applications \cite{dickey2005laser}, are obtained at distinct correlation-width values, $\delta|_{(\eta=1)} = 2{\thinspace}mm$, $\delta|_{(\eta=2)} = 1.45{\thinspace}mm$ and $\delta|_{(\eta=3)} = 1.2{\thinspace}mm$. This means that the donut shaped intensity profile evolves slowly towards a Gaussian profile for higher PHI beams. On reducing $\delta$, the outer ring transfers energy to the central core resulting in an increase in the on-axis intensity. For a higher PHI beam carrying a bigger dark-core region, the correlation in the fields has to be reduced to a larger extent to obtain a flat-top beam profile. This also indicates that the contribution of the unpolarized part is more in the flat-top higher PHI-PCVVB. Interestingly, as seen in Fig. \ref{fig:4}, the polarization distribution and Stokes phase map remain invariant with the change in $\delta$ ($\delta=3.8{\thinspace}mm,{\thinspace} 2.2{\thinspace}mm$). Intuitively, the change in coherence in the source plane would affect the SOP distribution, but owing to the basic property of the Fresnel diffraction integral, angular harmonic modulations are preserved \cite{gori2008partially}.  This physically means that the spread of each element of CSD matrix is identical for a fixed $\delta$ \cite{korotkova2005changes}. \textcolor{black}{Moreover, masking of a single radially polarized beam yields radially polarized lattice fields \cite{liang2017vector, pal2017generation}}.
\begin{figure*} [ht!]
\centering
\includegraphics[width=0.95\linewidth]{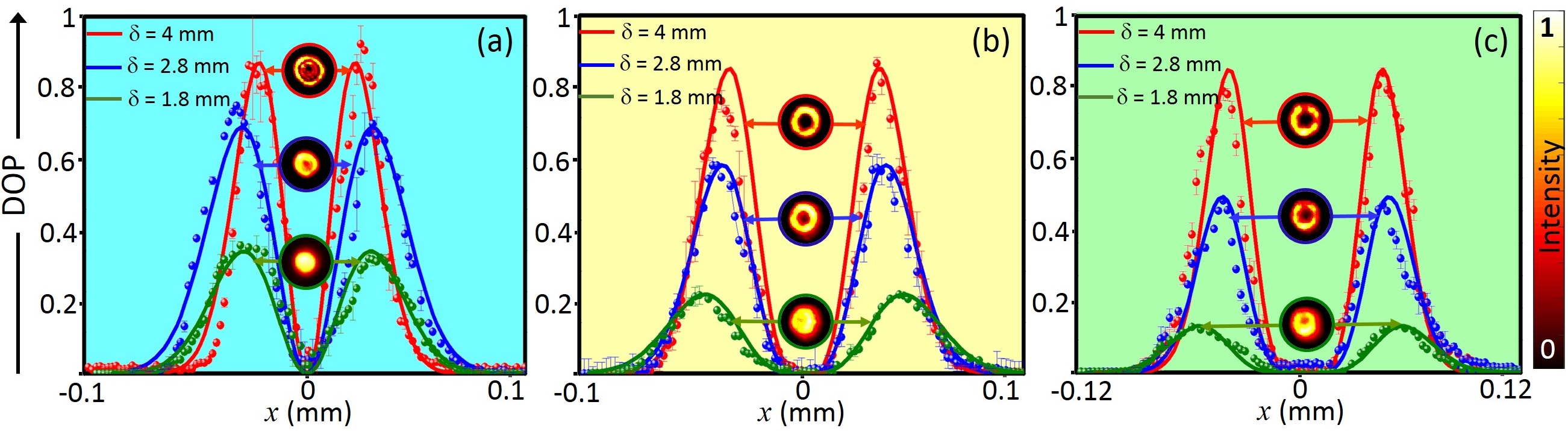}
\caption{Transverse DOP profile with varying correlation-width for (a) $|\eta|=1$, (b) $|\eta|=2$ and (c) $|\eta|=3$. Solid lines  are theoretically predicted DOP profiles and filled circles are experimentally obtained data-points. Inset shows the respective intensity profiles.}
\label{fig:5}
\end{figure*}
\begin{figure} [ht!]
\centering
\includegraphics[width=0.92\linewidth]{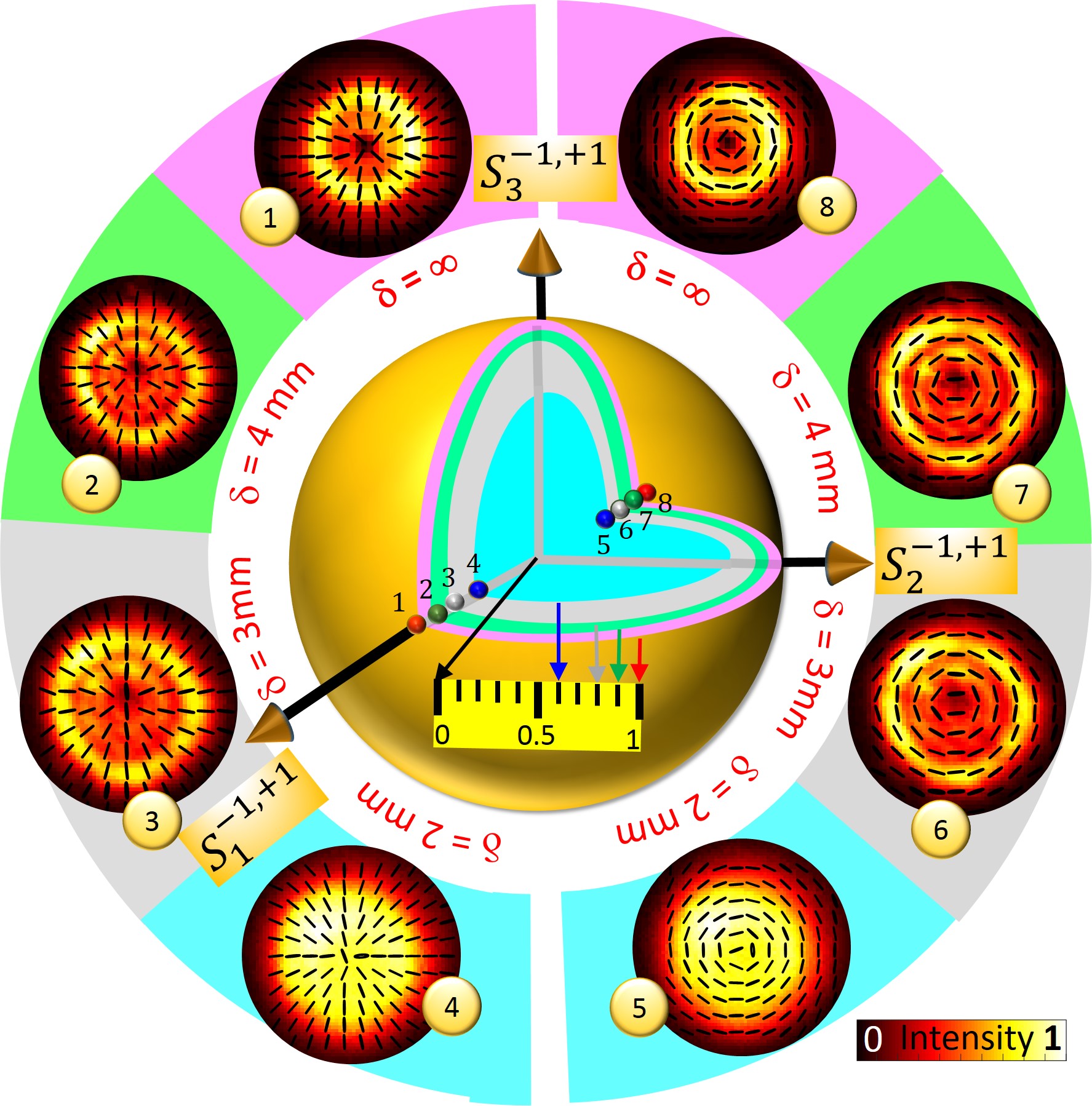}
\caption{Higher-order Poincar\'e sphere representing experimentally obtained unit-index PCVVBs. Orthogonal pairs (radial and azimuthal) are represented by diametrically opposite equatorial points. Points 1 to 4 (8 to 5) shows the beam-location as a function of correlation-width.}
\label{fig:6}
\end{figure}
Figure \ref{fig:5} illustrates the transverse DOP profile of the PCVVB of $|\eta|=1, 2 \thinspace\text{and}\thinspace 3$, for different correlation-widths ($\delta$). The on-axis DOP is zero and it increases symmetrically in the transverse direction. The maximum values of DOP for $|\eta|=1$ is higher than that obtained for $|\eta=2|$ and $|\eta=3|$. Also, the DOP-dip becomes wider on reducing $\delta$ or increasing $\eta$. Zero-DOP at the centre is due to the absence of polarized light field. The wider-dip for higher PHI-PCVVB implies a larger contribution of unpolarized content to the beam for a fixed $\delta$.

The intensity profile change due to the enhanced unpolarized part in an inhomogeneously polarized beam can be conceived by using a sphere similar to a higher-order Poincar\'e sphere (HOPS) \cite{milione2011higher}. The experimentally obtained beam profiles of unit index PCVVBs are shown on the HOPS in Fig. \ref{fig:6}. The radially (0,0) and azimuthally ($\pi$,0) polarized light beams are located diametrically opposite along the equatorial plane either on the outer/inner sphere. The surface of the HOPS represents highly coherent completely polarized light-field and the inner spherical shells correspond to the maximum DOP values at the edge of the respective beams. The radius of these spherical shells depends upon $\delta$. As we move from the periphery to the interior (Point 1 to 4 or Point 8 to 5), the unpolarized part increases and the DOP reduces. Similarly, one can construct HOPS of higher-index PCVVBs. The rate of coherence-induced depolarization effects is different for various orders of PHI beams and is implicitly reflected in the radial direction of the respective HOPS. 

In conclusion, coherence-induced polarization effects in various PHI-VVBs have been demonstrated. The irradiance profile and the DOP-dip are functions of both the correlation-width and the PHI of the beam. The prominent depolarization with decreasing transverse correlation-width in these PCVVBs enables beam profile shaping while preserving the SOP distribution; and provides an additional degree of freedom $-$ $"$DOP of VVB$"$ for a myriad of applications in the classical and quantum domain.
\section*{Acknowledgement}
SJ and SNK acknowledge IIT-Delhi post-doctoral fellowship.
\section*{Disclosures} 
The authors declare no conflicts of interest.
\bibliography{ppvb}

\begin{thebibliography}{10}
\newcommand{\enquote}[1]{``#1''}

\bibitem{wolf2007introduction}
E.~Wolf, \emph{Introduction to the Theory of Coherence and Polarization of
  Light} (Cambridge University Press, 2007).

\bibitem{wolf2003correlation}
E.~Wolf, {\protect\JournalTitle{Optics Letters}} \textbf{28}, 1078 (2003).

\bibitem{james1994change}
D.~F. James, {\protect\JournalTitle{J. Opt. Soc. Am. A}} \textbf{11}, 1641
  (1994).

\bibitem{korotkova2005changes}
O.~Korotkova and E.~Wolf, {\protect\JournalTitle{Optics Communications}}
  \textbf{246}, 35 (2005).

\bibitem{zhao2018controlling}
X.~Zhao, T.~D. Visser, and G.~P. Agrawal, {\protect\JournalTitle{Opt. Lett.}}
  \textbf{43}, 2344 (2018).

\bibitem{zhao2018degree}
X.~Zhao, T.~D. Visser, and G.~P. Agrawal, {\protect\JournalTitle{J. Opt. Soc.
  Am. A}} \textbf{35}, 1518 (2018).

\bibitem{zeng2020partially}
J.~Zeng, C.~Liang, H.~Wang, F.~Wang, C.~Zhao, G.~Gbur, and Y.~Cai,
  {\protect\JournalTitle{Optics Express}} \textbf{28}, 11493 (2020).

\bibitem{liu2019review}
X.~Liu, J.~Zeng, and Y.~Cai, {\protect\JournalTitle{Advances in Physics: X}}
  \textbf{4}, 1626766 (2019).

\bibitem{ji2009changes}
X.~Ji and X.~Chen, {\protect\JournalTitle{Opt. Laser Technol.}} \textbf{41},
  165 (2009).

\bibitem{kanseri2013optical}
B.~Kanseri, \emph{Optical Coherence and Polarization: An Experimental Outlook}
  (Lambert Academic, 2013).

\bibitem{Senthil}
P.~Senthilkumaran, \emph{Singularities in Physics and Engineering} (IOP
  Publishing, 2018).

\bibitem{freund2001polarization}
I.~Freund, {\protect\JournalTitle{Opt. Commun.}} \textbf{199}, 47 (2001).

\bibitem{dennis2009singular}
M.~R. Dennis, K.~O'Holleran, and M.~J. Padgett, \enquote{Singular
  \uppercase{O}ptics: \uppercase{O}ptical \uppercase{V}ortices and
  \uppercase{P}olarization \uppercase{S}ingularities,} in \emph{Prog. Opt.},
  (Elsevier, 2009).

\bibitem{milione2011higher}
G.~Milione, H.~Sztul, D.~Nolan, and R.~Alfano, {\protect\JournalTitle{Phys.
  Rev. Lett.}} \textbf{107}, 053601 (2011).

\bibitem{khan2020perturbation}
S.~N. Khan, S.~Deepa, G.~Arora, and P.~Senthilkumaran,
  {\protect\JournalTitle{J. Opt. Soc. Am. B}} \textbf{37}, 1577 (2020).

\bibitem{guo2011intensity}
L.~Guo, Z.~Tang, C.~Liang, and Z.~Tan, {\protect\JournalTitle{Opt. Laser
  Technol.}} \textbf{43}, 895 (2011).

\bibitem{wang2012experimental}
F.~Wang, Y.~Cai, Y.~Dong, and O.~Korotkova, {\protect\JournalTitle{Appl. Phys.
  Lett.}} \textbf{100}, 051108 (2012).

\bibitem{dong2012effect}
Y.~Dong, F.~Wang, C.~Zhao, and Y.~Cai, {\protect\JournalTitle{Physical Review
  A}} \textbf{86}, 013840 (2012).

\bibitem{gbur2020singularities}
W.~S. Raburn and G.~Gbur, {\protect\JournalTitle{Frontiers in Physics}}
  \textbf{8}, 168 (2020).

\bibitem{chen2014generation}
Y.~Chen, F.~Wang, L.~Liu, C.~Zhao, Y.~Cai, and O.~Korotkova,
  {\protect\JournalTitle{Phys. Rev. A}} \textbf{89}, 013801 (2014).

\bibitem{korotkova2008scintillation}
O.~Korotkova, {\protect\JournalTitle{Optics Communications}} \textbf{281}, 2342
  (2008).

\bibitem{ndagano2017characterizing}
B.~Ndagano, B.~Perez-Garcia, F.~S. Roux, M.~McLaren, C.~Rosales-Guzman,
  Y.~Zhang, O.~Mouane, R.~I. Hernandez-Aranda, T.~Konrad, and A.~Forbes,
  {\protect\JournalTitle{Nature Physics}} \textbf{13}, 397 (2017).

\bibitem{ram2017probing}
B.~S.~B. Ram, A.~Sharma, and P.~Senthilkumaran, {\protect\JournalTitle{Opt.
  Lett.}} \textbf{42}, 3570 (2017).

\bibitem{kanseri2010experimental}
B.~Kanseri and H.~C. Kandpal, {\protect\JournalTitle{Optics Express}}
  \textbf{18}, 11838 (2010).

\bibitem{lin2002tensor}
Q.~Lin and Y.~Cai, {\protect\JournalTitle{Opt. Lett.}} \textbf{27}, 216 (2002).

\bibitem{born2013principles}
M.~Born and E.~Wolf, \emph{Principles of Optics: Electromagnetic Theory of
  Propagation, Interference and Diffraction of Light} (Elsevier, 2013).

\bibitem{chen2014experimental}
Y.~Chen, F.~Wang, C.~Zhao, and Y.~Cai, {\protect\JournalTitle{Opt. Express}}
  \textbf{22}, 5826 (2014).

\bibitem{foley1991effect}
J.~T. Foley, {\protect\JournalTitle{J. Opt. Soc. Am. A}} \textbf{8}, 1099
  (1991).

\bibitem{dickey2005laser}
F.~M. Dickey, S.~C. Holswade, T.~E. Lizotte, and D.~L. Shealy, \emph{Laser Beam
  Shaping Applications} (CRC Press, 2005).

\bibitem{gori2008partially}
F.~Gori, {\protect\JournalTitle{Opt. Lett.}} \textbf{33}, 2818 (2008).

\bibitem{liang2017vector}
C.~Liang, C.~Mi, F.~Wang, C.~Zhao, Y.~Cai, and S.~A. Ponomarenko,
  {\protect\JournalTitle{Opt. Express}} \textbf{25}, 9872 (2017).

\bibitem{pal2017generation}
Ruchi, S.~K. Pal, and P.~Senthilkumaran, {\protect\JournalTitle{Optics
  Express}} \textbf{25}, 19326 (2017).

\end{thebibliography}

\bibliographyfullrefs{sample}


\ifthenelse{\equal{\journalref}{aop}}{%
\section*{Author Biographies}
\begingroup
\setlength\intextsep{0pt}
\begin{minipage}[t][6.3cm][t]{1.0\textwidth} 
  \begin{wrapfigure}{L}{0.25\textwidth}
    \includegraphics[width=0.25\textwidth]{john_smith.eps}
  \end{wrapfigure}
  \noindent
  {\bfseries John Smith} received his BSc (Mathematics) in 2000 from The University of Maryland. His research interests include lasers and optics.
\end{minipage}
\begin{minipage}{1.0\textwidth}
  \begin{wrapfigure}{L}{0.25\textwidth}
    \includegraphics[width=0.25\textwidth]{alice_smith.eps}
  \end{wrapfigure}
  \noindent
  {\bfseries Alice Smith} also received her BSc (Mathematics) in 2000 from The University of Maryland. Her research interests also include lasers and optics.
\end{minipage}
\endgroup
}{}
\end{document}